\documentstyle[prl,twocolumn,aps]{revtex}
\begin{document}
\input{epsf.tex} 
\epsfverbosetrue

\draft

\twocolumn[\hsize\textwidth\columnwidth\hsize\csname @twocolumnfalse\endcsname

\title{Existence and stability of atomic-molecular Bose-Einstein condensates}

\author{Benedict J. Cusack, Tristram J. Alexander,  Elena A. Ostrovskaya, 
and Yuri S.  Kivshar}

\address{Optical Sciences Centre, The Australian National University,
  Canberra ACT 0200, Australia}

\date{\today}

\maketitle

\begin{abstract}
  We analyze the model of an atomic Bose-Einstein condensate (BEC) parametrically coupled to a molecular BEC via a photoassociation process. We show that an interplay of nonlinear inter- and intra-species interactions leads to the formation of {\em mutually trapped states of a hybrid condensate}, which are spatially localized {\em even without a trap}. The untrapped atomic-molecular condensates with a large number of atoms in either fraction are shown to be {\em dynamically stable} in the framework of a mean field theory. 
\end{abstract}

]

\narrowtext

Recent progress in the field of the Bose-Einstein condensation is associated with the possibility of creating multi-component condensates of different, coherently coupled, atomic species. Furthermore, the possibility of creating a hybrid condensate of atomic and molecular fractions has been suggested \cite{am_theory},  and recently received some experimental support \cite{wynar}. In such an atomic-molecular condensate, atoms would bind into molecules coherently, through a (reversible) photoassociation-induced stimulated recombination in two-body collisions. 

The problem of generation of coupled atomic and molecular condensates  is fascinating from different points of view (see, e.g.,  Refs. \cite{wynar,heinzen,javanainen,kostrun}). Apart from the prospect of observing the first {\em molecular condensate}, this problem offers a unique atom-optics analogue to the process of second-harmonic generation in nonlinear optics.  The nonlinearity produced by coherent atomic-molecular coupling is expected to be responsible for a number of dynamical effects, including inter-species Josephson-like population oscillations \cite{heinzen,timmermans,salgueiro}, and ``clumping'' of the condensate due to the atom-optical analogue of modulational instability \cite{javanainen,kostrun}.  In a yet to be realized atomic-molecular BEC (AMBEC), the atom-molecular parametric coupling would compete with the atom-atom, molecule-molecule, and atom-molecule collisional interactions, which corresponds to the competing quadratic and cubic intensity-dependent response  in nonlinear optics \cite{optics}, and suggests the existence of liquid-like droplets of the atomic-molecular condensates \cite{timmermans}.

In this Letter, we present a comprehensive analysis of the mean-field model for an AMBEC created via a coherent stimulated Raman photoassociation process. Using the recently reported data for the intra- and inter-species scattering lengths \cite{wynar}, we find the existence domain for the stationary states of an AMBEC in an experimentally relevant parameter region. We then show that, under certain conditions, stationary states of an AMBEC are dynamically stable, which enables the creation of {\em self-confined two-species three-dimensional AMBEC droplets}, after the condensate is released from a confining trap. Untrapped droplets of the coherent matter waves resemble self-trapped spatial solitons supported by parametric interaction of optical waves \cite{optics}.

We study the dynamics of the atomic and molecular condensates in a spherically symmetric, three-dimensional trap, dynamically coupled by a free-bound Raman photoassociation process \cite{heinzen,timmermans}, in the framework of the coupled mean-field Gross-Pitaevskii (GP) equations for the macroscopic wave functions of the two species, 
\begin{equation}
\label{model_eq}
\begin{array}{l}
{\displaystyle i\frac{\partial \psi_a}{\partial t} + \frac{1}{2}\Delta\psi_a 
- \frac{1}{2}r^{2}\psi_a - \chi \psi_a^*\psi_m e^{i\delta t}}\\*[9pt]   
{\displaystyle - (U_{aa}|\psi_a|^2 + U_{am}|\psi_m|^2) \psi_a  = 0,}\\*[9pt]
{\displaystyle i\frac{\partial \psi_m}{\partial t} + \frac{1}{4}\Delta\psi_m 
- \Omega^2 r^{2}\psi_m - \frac{1}{2}\chi \psi_a^2 e^{-i\delta t} }\\ *[9pt] 
{\displaystyle - ( U_{mm} |\psi_m|^2 + U_{am}|\psi_a|^2)\psi_m = 0.}\\ *[9pt]
\end{array}
\end{equation}
Here the wave functions, time, and spatial coordinates are measured in the units of  $(\hbar/m\omega)^{-3/4}
$, $\omega^{-1}$, and $(\hbar/m\omega)^{1/2}$, respectively, where $\omega$ is the characteristic frequency of the trap for the atomic BEC fraction. The interaction strengths $U_{ij}$ are measured in the units of $(\hbar\omega)^{-1}(\hbar/m\omega)^{-3/2}$, and the parametric coupling strength, $\chi$, is measured in the units of $(\hbar\omega)^{-1}(\hbar/m\omega)^{-3/4}$. We also introduce the dimensionless Raman detuning parameter $\delta$, measured in the units of $\omega^{-1}$, and the ratio between the trap frequencies for the molecular and atomic condensate fractions, $\Omega=\omega_m/\omega$. 
\noindent \parbox[l]{8cm}{
\noindent  
\begin{table}
\caption{Model parameters for $^{87}$Rb AMBEC.}
\label{tabdata}
\begin{tabular}{|c|c|c|} 
\hline
\text{Model} & \text{Parameter} & \text{Dimensionless} \\ 
\text{parameters} & \text{values} & \text{values} \\
\tableline
$U_{aa}$ & $5.3\times10^{-51}\mathrm{Jm}^3$ \cite{heinzen} & $0.062$ \\
$U_{am}$ & $-6.9\times10^{-51}\mathrm{Jm}^3$ \cite{wynar}& $-0.084(\pm 0.07)$ \\
$U_{mm}$ & $9.9\times10^{-51}\mathrm{Jm}^3$ \cite{umm}& $0.12$ \\
$\chi$ & $7.4\times10^{-41}\mathrm{Jm}^{3/2}\cite{heinzen}$ & $1.09$ \\
\hline
\end{tabular}
 \end{table}
}
To specify the model parameters, we consider the case of $^{87}$Rb, for which the recent experiments on the formation of ultracold molecules were conducted \cite{wynar}. Realistic values for the parameters can be deduced from a number of available experimental data (see Table 1). We also take $\omega/2 \pi = 100$, and $\Omega = 1$. The atom-molecule two-body interaction strength was measured very recently \cite{wynar}. Despite large uncertainty in the value of the corresponding $s$-wave scattering length, $a_{am}=-180a_0\pm 150a_0$, it is apparent that, unlike the atom-atom interaction, the atom-molecule interaction is {\em attractive}. This fact, as we show below, results in {\em a competition of attractive and repulsive interactions} in an AMBEC.

%\section{Stationary states}
To find possible steady states, we look for radially-symmetric solutions of Eqs. (\ref{model_eq}) in the form: $\psi_a({\bf r},t)=\phi_a(r) \exp(-i \mu_a t)$, $\psi_m({\bf r},t)=\phi_m(r) \exp(-i\delta t-i \mu_m t +i\theta)$, where $\phi_a(r)$ and $\phi_m(r)$ are positively defined, real functions, $\mu_a$ and $\mu_m$ are the chemical potentials, and $\theta$ is a relative phase of the condensate fractions. The substitution of the above ansatz into Eqs. (\ref{model_eq}) reveals that, in order for a stationary AMBEC to exist, two conditions must be satisfied: $\mu_m=2\mu_a\equiv 2\mu$ and $\theta=\{0,\pi\}$. The first condition is reminiscent of the phase-matching of a fundamental and second-harmonics waves in an optical parametric process. The second condition implies that there are at least two {\em physically different} stationary configurations of the AMBEC, with the two components being {\em ``in-''} or {\em ``out-of-phase''}. The equations for $\phi_a(r)$ and $\phi_m(r)$ become:
\begin{equation}
\label{st_eq}
\begin{array}{l}
{\displaystyle \frac{1}{2}\left(\Delta_r 
- r^{2}\right)\phi_a - \sigma\chi \phi_a \phi_m +\mu\phi_a}\\ *[9pt]
{\displaystyle - (U_{aa}\phi_a^2 + U_{am}\phi_m^2) \phi_a  = 0,}\\ *[9pt]
{\displaystyle \frac{1}{4}\left(\Delta_r 
- 4r^{2}\right)\phi_m +\delta \phi_m - \sigma\frac{\chi}{2} \phi_a^2 +2\mu\phi_m} \\*[9pt] {\displaystyle - ( U_{mm} \phi_m^2 + U_{am}\phi_a^2)\phi_m = 0,}\\*[9pt]
\end{array}
\end{equation} 
where $\Delta_r=d^2/dr^2+(2/r)d/dr$ and $\sigma\equiv \cos~\theta=\pm 1$. In what follows, we assume near-resonant parametric coupling and set $\delta=0$. 

From Eq. (\ref{st_eq}) it can be seen that the intra- and inter-species interactions act as repulsive or attractive potentials for the condensate components. Without the parametric (Raman) coupling, the net effect of the cubic interactions is determined by the sign of $\Delta U\equiv U_{aa}U_{mm}-U^2_{am}$ \cite{timmermans}. Since the atom-molecule interaction is attractive, the net-attractive and net-repulsive regimes occur when $\Delta U<0$ and $\Delta U>0$, respectively. In addition, the quadratic Raman coupling is effectively {\em repulsive} for in-phase components ($\sigma=+1$), and {\em attractive} for out-of-phase components ($\sigma=-1$). Depending on the relative phase, $\theta$, the system admits two types of stationary AMBEC states. Typical families of stationary states in net-repulsive regime are shown in Fig. 1; we identify them by the mnemonics $am_+$ (for $\sigma=+1$) or $am_-$ (for $\sigma=-1$).

The $am_+$ states are dominated by the repulsive cubic and parametric interactions, and exist only in the presence of a trap. Since the system (\ref{st_eq}) admits stationary solutions in the form of a {\em pure molecular condensate}, the family of the hybrid solutions branches off the pure molecular state at a bifurcation point (upper open circle in Fig. 1). In the vicinity of this point the atomic fraction is small. The physical meaning of this bifurcation and the coexistence of the pure molecular and AMBEC solution families is the following: unless some of the atomic condensate is present, the two-component AMBEC will not form again, if a complete transformation into  the molecular fraction has occured. This is a feature of the GP model which ignores spontaneous transitions from the molecular state (see discussions in \cite{hope}). 
\begin{figure}
\setlength{\epsfxsize}{6.5cm} \centerline{ \epsfbox{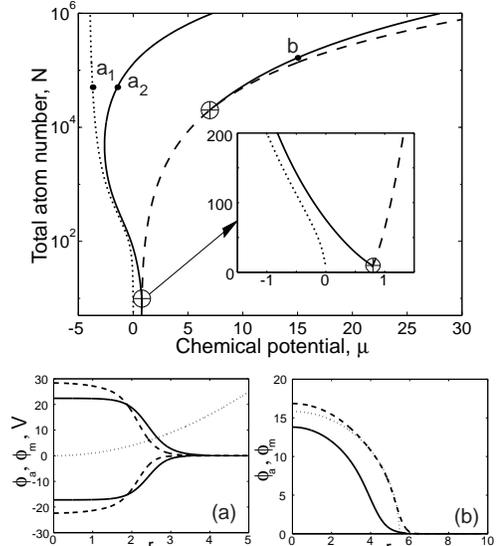}}
\caption{Top: Families of AMBEC stationary states at $U_{am}=-0.054$: trapped out-of-phase ($a_2$) and in-phase ($b$) states (solid), untrapped out-of-phase state ($a_1$, dotted) and pure molecular BEC state (dashed).  The trapped $am^{\rm t}_-$  solution (solid) branches off the molecular BEC state at the bifurcation point (inset).  (a) Radial wave functions, $\phi_a$ and $\phi_m e^{i\theta}$, of trapped $am^{\rm t}_-$ (dashed) and untrapped $am^{\rm ut}_-$ (solid) states for $\mu=-1.45$ and $\mu=-3.7$, respectively, shown with the trapping potential $V(r)=r^2$ (dotted); $N=5.04\times10^4$. (b) Radial wave functions of a trapped $am_+$ stationary state for $\mu=15$ and $N=1.66\times10^5$, shown with the Thomas-Fermi approximation for $\phi_m$ matching the number of atoms in the molecular fraction (dotted).}
 \label{fig1}
\end{figure} 
Remarkably, the shape of the molecular component remains close to that obtained by the Thomas-Fermi approximation even far from the bifurcation point, while the atomic component is moderate [see Fig. 1(b)]. This is due to the fact that, for the specific values of parameters corresponding to $^{87}$Rb case (see Table 1), the effective potential experienced by the molecular component is dominated almost entirely by the molecular self-interaction; the inter-species and Raman interactions, being of different sign and comparable magnitude, counteract and nearly cancel each other. Far from the bifurcation point, the density-dependent cubic interactions dominate over Raman coupling and the shape of the condensate deviates from that found by the Thomas-Fermi approximation. 

In the presence of a trap, the $am_-$ states branch off the
pure molecular state at a small positive $\mu$ (Fig. 1, inset) and exist for $\mu<0$. For $\Delta U \approx 0$, the cubic interactions nearly cancel each other so that the shapes of the condensate components are almost mirror images of each other, and the condensate is strongly affected only by the {\em attractive} Raman coupling. The resulting attractive interaction leads to {\em self-trapping of the two-component condensate}. Consequently, the $am_-$ states can exist {\em with} ($am^{\rm t}_-$) or {\em without} ($am^{\rm ut}_-$) the external trap. These states also exist when $\Delta U \neq 0$, and  the corresponding solutions for the net-repulsive regime are shown in Figs. 1(a,b). Even a subtle change in the magnitude of $\Delta U$ results in a dramatic change of the peak density of the stationary states.  For a fixed $\Delta U$ and increasing particle number, the $am^{\rm ut}_-$ solutions broaden spatially while keeping roughly the same peak density, asymptotically approaching the unbounded, homogeneous AMBEC. In contrast, the spread of the $am^{\rm t}_-$ states is hindered by the external trap, so they rapidly grow in the peak density. As a consequence, the two initially close families of $am_-$ depart at large $N$ (see Fig. 1). At the moderate values of $N$, while the families are still close, the condensate wave functions are weakly affected by the confining trap. 

\begin{figure}
\setlength{\epsfxsize}{8.5cm} \centerline{ \epsfbox{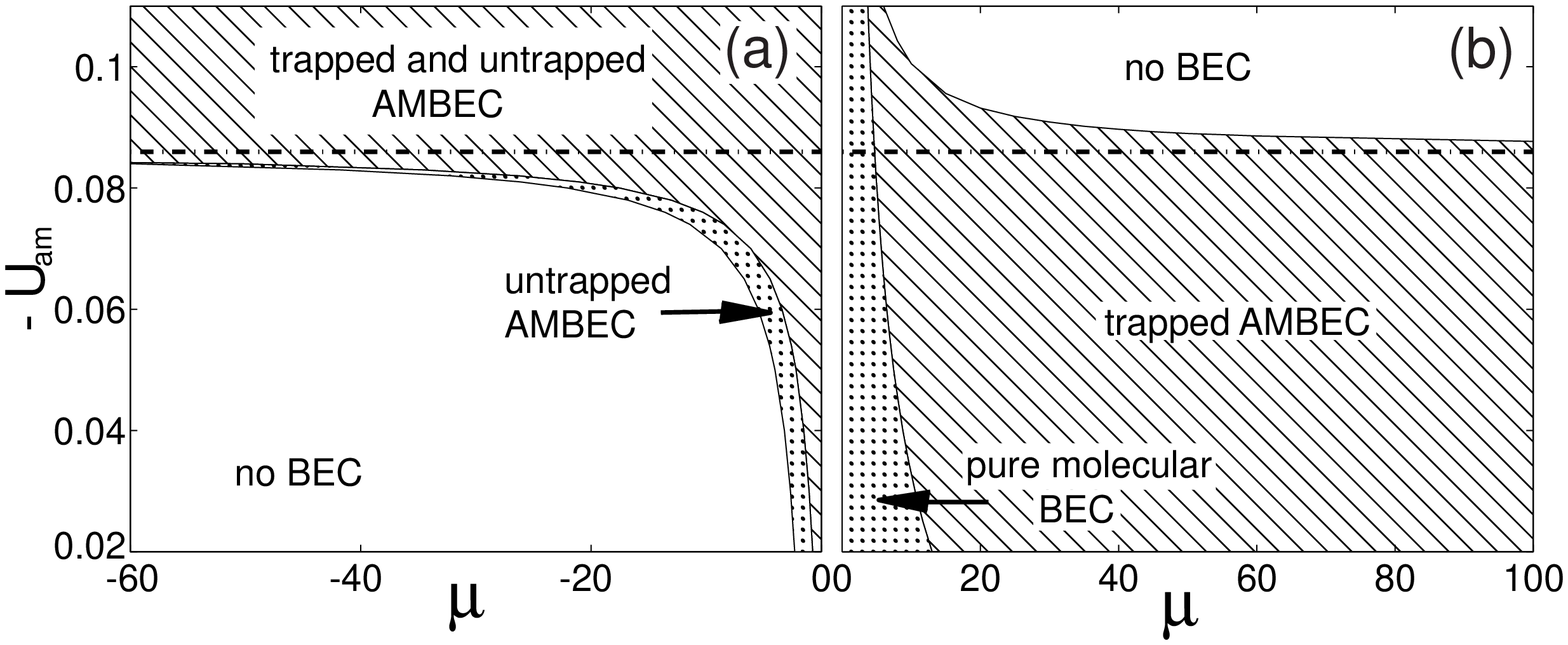}}
\caption{Existence domains in the parameter space $\{-U_{am}, \mu\}$.  (a) Trapped (hashed) and untrapped (dotted) stationary out-of-phase states.  (b) Trapped in-phase AMBEC (hashed) and trapped pure molecular BEC (dotted). Dash-dotted line marks the border $\Delta U=0$ between the net-attractive (above the line, $\Delta U<0$)  and net-repulsive (below the line, $\Delta U>0$) regimes.}
 \label{fig2}
\end{figure}
We note that the specific choice of parameters in Table 1 does not affect the generality of our results as the magnitude of $\Delta U$ only affects the densities of the stationary AMBEC fractions. However, the properties of the AMBEC stationary states in the net-attractive regime of cubic interactions differ drastically from those in the net-repulsive regime. The two different regimes can be entered by varying either of the interaction strengths $U_{ij}$ (e.g., by tuning $U_{aa}$ through a Feshbach resonance). Here we vary the $U_{am}$, within the large uncertainty in its value (see Table 1). The existence regions for both in-phase  and out-of-phase states are shown in Figs. 2(b) and 2(a), respectively.  For the $am_+$ states, the prevailing attractive interactions above the line $\Delta U=0$ ramps up the density of the condensate at the center of trap, until the AMBEC states cease to exist at a certain $U_{am}(\mu)$ [see Fig. 2(b)]. The $am^{\rm t}_-$ states exist everywhere in the net-attractive domain [see Fig. 2(a)], and since both cubic and quadratic (parametric) interactions are effectively attractive, the AMBEC grows in central density, becoming highly localized with growing $|\mu|$. Both $am_+$ and $am^{\rm t}_-$ states in a net-attractive regime are likely to be unstable against collapse.

%\section{Stability}
The AMBEC states in the net-repulsive regime are of great interest from the experimental point of view, provided they are stable. A linear stability analysis for the homogeneous states, the amplidudes of which are asymptotically approached by the $am^{\rm ut}_-$ and $am^{\rm t}_-$ states at the large particle number, show that both these states can be stable. Furthermore, our numerical analysis of the dynamical stability of AMBEC states performed by solving the system (\ref{model_eq}) with the initial conditions from different regions of the existence domain (see Fig. 2) confirmed that the $am_-$ states are likely to be {\em stable}, whereas all $am_+$ states are {\em dynamically unstable}.
%\section{Dynamics}
\vspace{-4mm}
\begin{figure}
\vspace{-4mm}
  \setlength{\epsfxsize}{7.5cm} \centerline{ \epsfbox{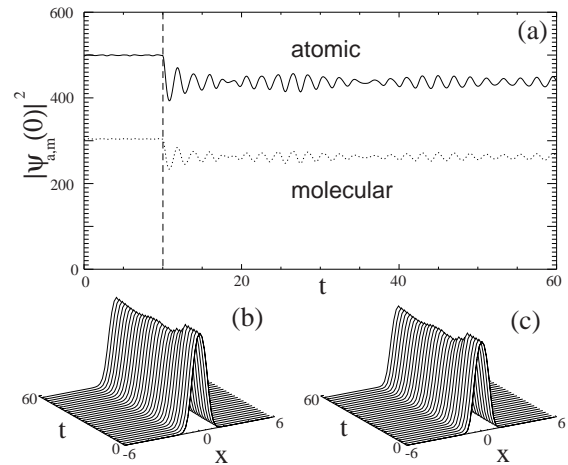}}
  \caption{
    (a) Evolution of an initially trapped, dynamically stable, stationary state $am^{\rm t}_-$, for $\mu = -2.5$ and $U_{am} = -0.054$, released from the trap at $t=10$ (dashed). (b,c) Evolution of the atomic and molecular components. Parameters for the initial state are as in Table 1.}
  \label{fig3}
\end{figure}
Technically, the formation of an AMBEC should begin within a trap. However, the co-existence, in the same parameter region, of {\em two types of dynamically stable states}, $am^{\rm t}_-$ and $am^{\rm ut}_-$, is a key to creation of AMBEC droplets without a trap. Namely, an $am^{\rm t}_-$ state can be created, and then the trap can be turned off, which can cause a transition to the dynamically stable $am^{\rm ut}_-$ state, provided the shapes and number of atoms of the condensates in the two states are close. This can be achieved by tuning the parameters so that the respective stationary states are weakly affected by the trapping potential (see Fig. 1). An example of the typical dynamics in this case is shown in Fig. 3, where the initially trapped out-of-phase state displays stable behavior. After the trap is turned off, the AMBEC quickly settles into small-amplitude oscillations around a new (untrapped) stable stationary state.

A natural route to the formation of a hybrid state is generation of the molecular component from a pure atomic condensate. Due to the parametric coupling, there is no stationary state corresponding to a pure atomic fraction, so formation of the molecular fraction starts immediately \cite{heinzen}. The initial stages of the AMBEC condensate formation are shown in Fig. 4(a). Without any dissipative processes, it is hard to anticipate that a dynamically stable stationary state would be created in a trap. However, even in this case the unstable dynamics of the condensate {\em can be suppressed by turning the trap off}. Figures 4(a-c) demonstrate that, after initial transitional stage, the untrapped condensate does settle into oscillations around a stable state. The transition, however, takes much longer than in the case of release of an already formed stationary $am^{\rm t}_-$ condensate (cf. Fig 3).
 \begin{figure}
  \setlength{\epsfxsize}{7.5cm} \centerline{ \epsfbox{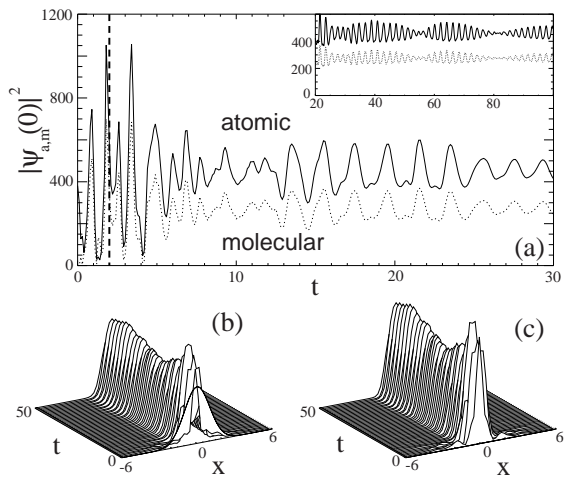}}
  \caption{
    (a) Generation of a trapped AMBEC from an atomic fraction (Gaussian initial profile) with subsequent release from the trap at $t=2$ (dashed line). Inset shows the long-term evolution. (b,c) Evolution of the atomic and molecular components. Parameters are as in Fig. 3}
  \label{fig4}
\end{figure}
It is important to mention that, in order to understand how a realistic 
AMBEC will relax to one of the stationary states,  one should account for the various effects of {\em losses}.  This can be done by adding phenomenological dissipative terms into the coupled GP equations (see also Refs. \cite{heinzen,timmermans}). Additionally, one could take into account the effect of {\em quantum fluctuations}, by going beyond the two-mode approximation of the GP equations \cite{holland,goral}, and deriving {\em the stochastic field equations} for the full quantum states of the condensates that include the GP equations as the noiseless limit \cite{hope}.  The resulting effective damping of the coherent atomic-molecular transfer oscillations will modify the process of formation of the AMBEC (especially important after a near total conversion to the molecular fraction \cite{hope}). However, all these effects will not significantly affect the dynamically stable stationary states, once and if, the system has settled into one of them, and they can be easily accounted for in the framework of the adiabatic approximation.   

In conclusion, we have analyzed the mean-field model of an atomic-molecular condensate and identified the existence domains for its stationary states. Our main finding is that, depending on the relative phase between the molecular and atomic fractions, two types of stationary states can be created in a confining trap. These are:  (i) Thomas-Fermi-like states, existing due to a balance of the trapping potential and nonlinear interactions, and (ii) self-trapped states, dominated by the (attractive) parametric inter-species interactions. In a certain region of the existence domain, the latter states are only weakly affected by the trap, and, upon release from the trap, they can be transformed into dynamically stable AMBEC droplets that are localized even without a trap. We believe that this result is potentially useful for the atom (molecular) laser applications where the formation of stable untrapped coherent matter waves is needed.

The authors are indebted to P.D. Drummond, J. Hope, N. Robins, and C. Savage for useful discussions.


\begin{references}
\bibitem{umm} Since the molecular condensate has not been experimentally obtained yet, the order of magnitude of the molecule self-interaction strength ($U_{mm}$) is estimated by treating the cold molecular fraction created in the experiment \cite{wynar} as a single-component molecular BEC with the temperature less or $\sim 10^2$ nK and a number of molecules $\sim 10^5$. The estimated (with the help of the Thomas-Fermi approximation) scattering length $a_{mm}$ is less or $ \sim 10^3 a_0$, which is comparable to the values of $a_{aa}$ and $a_{am}$ and therefore cannot be ignored. We chose the value of $U_{mm}$ so that $\Delta U$ is close to zero for the mean value of $U_{am}$.
\end{references}
\end{document}